\documentstyle[11pt,newpasp,twoside,psfig]{article}
\index{supernova remnants}
\index{X-ray}

\def\edcomment#1{\iffalse\marginpar{\raggedright\sl#1\/}\else\relax\fi}
\reversemarginpar

\newcommand{\subj}{MSH14-6{\it 3}}
\newcommand{\xmm}{XMM-Newton}
\newcommand{\Te}{$kT_{\rm e}$}
\newcommand{\net}{$n_{\rm e}t$}

\newcommand{\spex}{{\it SPEX}}

\newcommand{\mekal}{{\it mekal}}

\newcommand{\jetp}{JETP}

\begin{document}
\title{XMM-Newton observations of \subj\ (RCW 86)}
\author{Jacco Vink}
\affil{Chandra fellow,
Columbia Astrophysics Laboratory,  MC 5247,
550 West 120th St., New York, NY 10025}
\author{Johan Bleeker, Jelle Kaastra, Kurt van der Heyden}
\affil{SRON, %National Institute for Space Research, 
Sorbonnelaan 2, NL-3584 CA, Utrecht, The Netherlands}
\author{John Dickel}
\affil{Astronomy Department, University of Illinois, 1002 West Green Street, 
Urbana, IL 61801, USA}

\begin{abstract}
We present an analysis of the X-ray emission of the supernova remnant \subj, 
which was partially covered by three observations with \xmm.
The detection of Fe K emission at 6.4~keV, and the lack of
spatial correlation between hard X-ray and radio emission is evidence
against a dominant X-ray synchrotron component. We argue that
the hard X-ray continuum is best explained by 
non-thermal bremsstrahlung from a supra-thermal tail to an otherwise cool 
electron gas. 
The existence of low electron temperatures
is supported by low temperatures found in other parts of the remnant,
which are as low as $0.2$~keV in some regions.
\end{abstract}

\section{Introduction}
The X-ray emission from the supernova remnant \subj\ (RCW 86, G315.2-2.3) 
is characterized by spatially distinct soft and hard X-ray components
(Vink, Kaastra, \& Bleeker 1997). The soft X-ray emission has a thermal nature,
but the hard X-ray emission shows relatively little line emission, suggesting
X-ray synchrotron radiation (Borkowski et al. 2001), 
similar to SN 1006 (Koyama et al. 1995). 
Borkowski et al. argue that additional evidence for X-ray synchrotron 
radiation is the
spatial correlation between hard X-ray and radio emission.

A problem is, however, that
the hard X-ray emission is accompanied by
Fe K line emission at 6.4~keV (Vink et al. 1997).
The energy of the line emission indicates that the emission is from 
underionized iron (Fe XVII or lower),
but it also proofs that the line emitting plasma contains 
electrons with energies in excess of 7.1~keV.
These electrons should give rise to bremsstrahlung emission,
suggesting that at least part of the hard X-ray continuum
is bremsstrahlung.

We present here \xmm\ data of \subj. 
\xmm\ offers a superior sensitivity and spatial resolution 
(FWHM $\sim$ 6\arcsec) compared to ASCA.
This allows for a better separation of the hard and soft X-ray emitting 
regions.
The complete set of observations covers most of the remnant,
but here we only discuss data from the southeastern,
central and northwestern part, which were observed
in August 2000 with exposures of $\sim 12$~ks to $17$~ks.
More recent observations of the southwestern part 
will be presented in a future article.

\begin{figure}

\psfig{figure=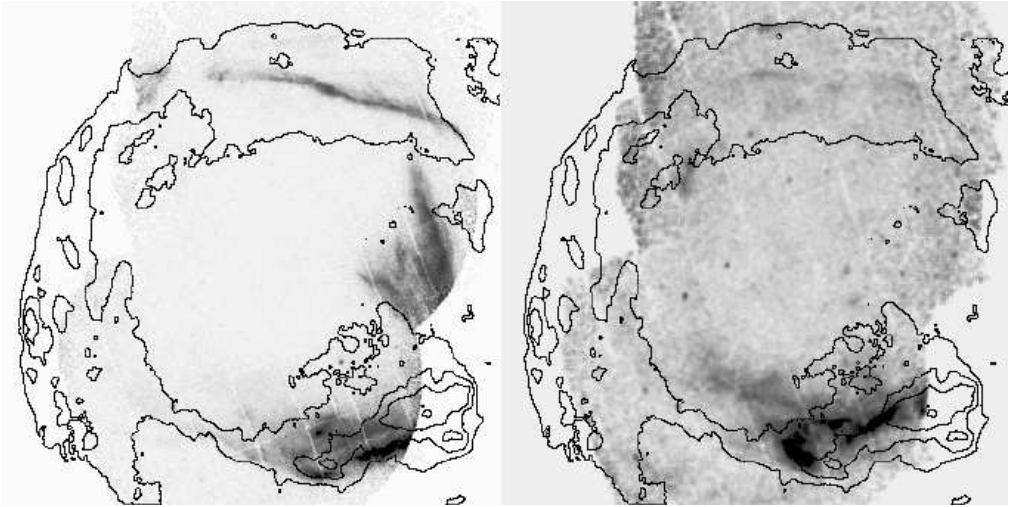,width=\textwidth}
\caption{Soft X-ray (0.5-1 keV), left, and hard X-ray (2-7~keV) \xmm\ mosaics
of \subj, combining exposure and effective area corrected data from all three CCD cameras; 
radio contours (ATCA, Dickel, Strom, \& Milne, 2000) are overlayed.
\label{mosaics}}
\end{figure}

\section{The soft, thermal X-ray emission}
\begin{figure}
\hbox{
\psfig{figure=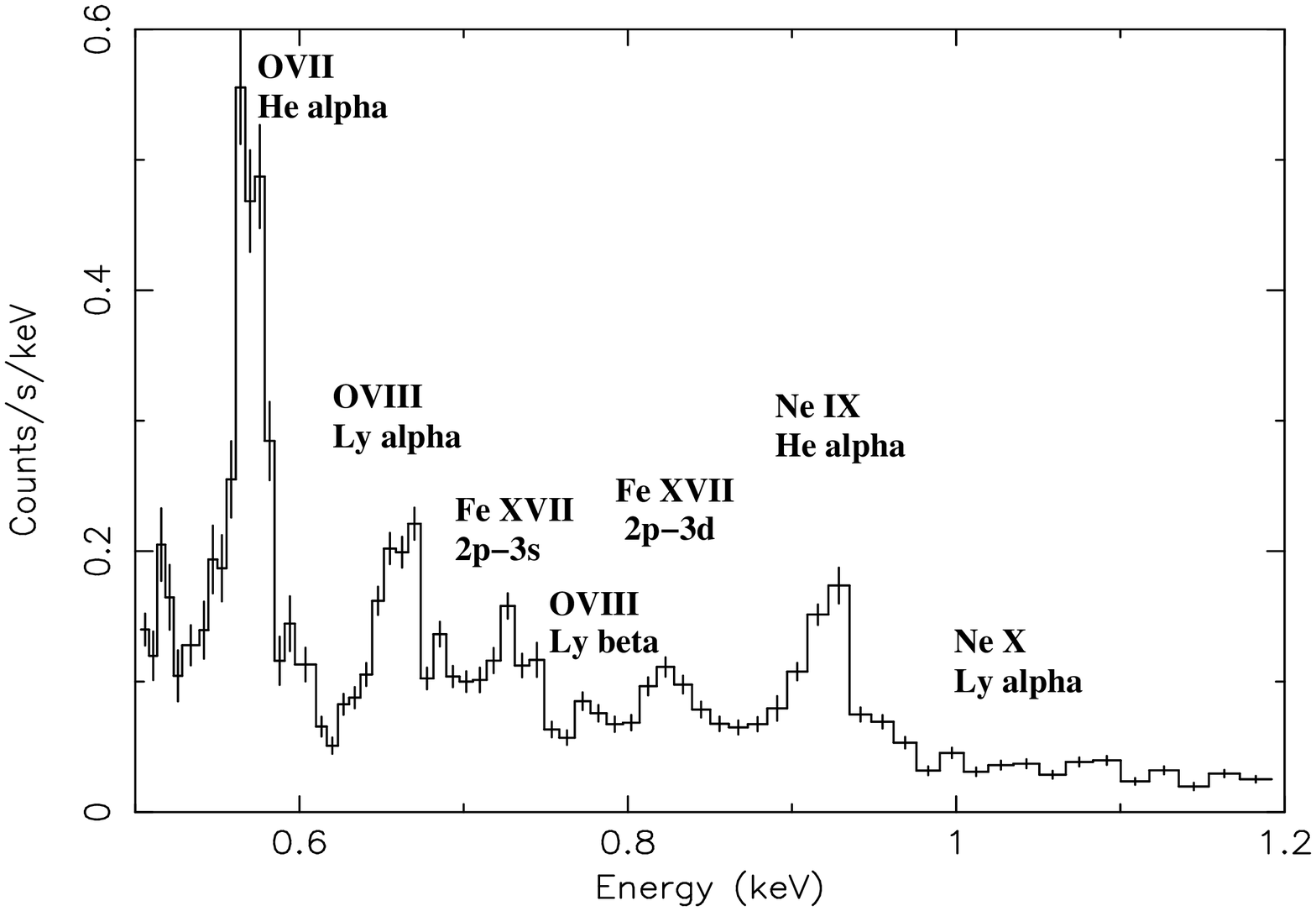,width=0.5\textwidth}
\psfig{figure=vinkj1_2b.ps,width=0.5\textwidth,angle=-90}}
\caption{
Left: 
high resolution RGS spectrum from the northwestern shell.
Right: EPIC-PN spectra from various regions of the remnant. 
The top spectrum is from a region north of the
northern shell; the other two spectra are from regions in the southeast.
\label{spectra}
}
\end{figure}

The soft X-ray emitting regions have a relatively low ionization characterized
by O VII, O VIII,  Ne IX and Fe XVII with an interesting 
variation in the relative importance of O VII emission.
The line emission is best illustrated by the spectrum from the northern shell
obtained with
the reflective grating spectrometer (RGS) (Fig. \ref{spectra}a).
The spectrum is dominated by O VII emission, as is also the case for regions
north of the shell and the fainter soft X-ray emitting regions in 
the southeast. Fig.~\ref{spectra}b shows the variation in O VII and O VIII 
emission, as observed by the EPIC PN instrument (Str\"uder et al. 2001).

The low ionization is likely to be the result of a low electron
temperature and a small ionization timescale. 
Unfortunately, these two plasma properties are not well constraint. 
For the region with the lowest ionization,
a region north of the northern shell, a temperature above 0.5~keV is only
possible if the ionization timescale is $\log($\net$) < 9.9$. 
A lower limit
to the temperature can be obtained by fitting a collisional ionization 
equilibrium (CIE) model, 
indicating an electron temperature possibly 
as low as 0.09~keV. The brightest soft X-ray emitting regions in the 
southeast may have \Te = 0.17~keV (CIE).\footnote{For the CIE model we used 
\mekal\ (Mewe et al. 1995),
for the NEI model we used the model contained in the \spex\ program 
which is based on \mekal.}
\enlargethispage{1mm}

The observed temperature range (0.1 - 0.5~keV) is lower than indicated by
the ASCA spectra (which was not very sensitive to O VII emission), but is 
consistent with the measured width of the H$\alpha$\ and  
H$\beta$ lines, which, under assumption of full electron-ion equilibration,
imply shock velocities of  310 - 605~km/s (Ghavamian et al. 2001).

\section{The nature of the hard X-ray emission}
If it were not for the
presence of the Fe K emission, the most likely mechanism for the X-ray continuum would be synchrotron radiation.
The presence of Fe K emission, which is confirmed by the \xmm\ data 
suggests, however, that at least part of the 
hard X-ray emission is bremsstrahlung.

The \xmm\ data indicates an Fe K equivalent width of $\sim 0.25$~keV. Modeling
bremsstrahlung and Fe K line emission from thermal and non-thermal electron 
distributions using the well known Lotz formula (e.g. Mewe 1999),
we find that for the observed equivalent width and solar abundances,
a temperature of at least 3~keV or a  electron power law index
$\geq -2.5$ are required, the latter value is
consistent with the observed power law index of $-2.7\pm0.1$.

Emission from underionized Fe suggests
that other elements may be underionized as well, but are not observed in
the spectra as lower Z elements have a smaller probability
for a radiative transition after inner shell ionization.
Vink et al. (1997) have argued that 
non-thermal bremsstrahlung may be a more appropriate model 
than thermal bremsstrahlung, 
as the cool, thermalized, part  
of the non-thermal distribution delays the
ionization of the plasma.
Moreover,
the low temperatures and shock velocities from other parts of the 
remnant are in sharp contrast to the $>3$~keV temperatures needed to explain
the Fe K emission.
Non-thermal distributions have been predicted for heating and acceleration of 
particles by
collisionless shocks (Bykov \& Uvarov).
There are, however, some problems with this interpretation. For instance,
the lack of O VII emission from the hard X-ray emitting regions require a low
temperature and additional non-equilibration ionization effects.

The statistics for the Fe K emission is limited, but its
spatial distribution is consistent with that of the hard X-ray
continuum, suggesting the two are related. This is hard to reconcile with
a synchrotron model, for which the continuum is caused by electrons
with energies in excess of 10~TeV, and the line emission is caused by electrons
with energies lower by 9 orders of magnitude.

With the better spatial resolution and sensitivity of
\xmm\ it is now also clear that, even in a qualitative sense,
there is not a good correlation
between hard X-ray and radio synchrotron emission, 
which was one of the principle arguments 
for synchrotron emission put forward by Borkowski et al. (2001), based
on the poor resolution ASCA hard X-ray maps.
Fig.~\ref{mosaics} shows this in particular for the 
southeastern region, where the hard X-ray emission bends towards the center, 
where there is no obvious radio counterpart.

\section{Concluding remarks}
Based on \xmm\ observations we have argued that the hard X-ray emission 
from \subj\ is (non-thermal)
bremsstrahlung rather than synchrotron emission. 
The mysterious X-ray emission may play a key role in studying
non-thermal processes associated with collisionless shock heating and
acceleration.
To understand the nature of the hard X-ray
emission, detailed maps of the continuum and Fe K emission are needed.
A close correlation between the two would indicate a bremsstrahlung
nature. A lack of correlation would indicate that the
hard X-ray emission comes from two distinct components:
ultrarelativistic electrons and a hot plasma.
This issue can therefore be resolved by a deep \xmm\ observation.

\acknowledgments
J. Vink is supported by the NASA 
through Chandra Postdoctoral Fellowship Award number PF0-10011
issued by the Chandra X-ray Observatory Center.%, which is operated by the
%Smithsonian Astrophysical Observatory for NASA.% under contract NAS8-39073.


\begin{references}
\reference Borkowski, K.J., Rho J., Reynolds, S.P., \& Dyer, K. 2001, \apj, 550, 334
\reference Bykov, A.M., \& Uvarov, Y.A., 1999, \jetp, 88, 465
\reference Dickel, J.R., Strom, R.G., \& Milne, D.K. 2000, \apj, 546, 447
\reference Ghavamian, P., Raymond, J., Smith, R.C., \& Hartigan, P. 2001, 
	\apj, 547, 995
\reference Koyama, K., et al. 1995, \nat, 378, 255
\reference Mewe, R., Kaastra, J., \& Liedahl, D.A. 1995,  Legacy 6, 16
\reference Mewe, R. 1999, in X-ray Spectroscopy in Astrophysics, 
        ed. J. van Paradijs and J. Bleeker (Springer-Verlag)
\reference Str\"uder, L., et al. 2001, \aap, 365, L18
\reference Vink, J., Kaastra J. , \& Bleeker, J. 1997, \aap, 328, 628
\end{references}
\end{document}